\documentclass{aa} 

\usepackage{graphics}

\begin{document}

\title{Polarimetry of Li-rich giants
\thanks{Based on observations obtained at the {\it Observat\'orio do Pico dos Dias}, LNA/MCT, Itajub\' a, Brazil.}}

\author{ A. Pereyra\inst{1}
\and B. V. Castilho\inst{2}
\and A. M. Magalh\~aes\inst{1} }

\offprints{A. Pereyra,
\email{antonio@astro.iag.usp.br}}

\institute{
Departamento de Astronomia, IAG, Universidade de S\~ao Paulo, Rua do Mat\~ao 1226, S\~ao Paulo, 05508-900, Brazil
\and
Laborat\'orio Nacional de Astrof\'{\i}sica/MCT, CP 21, Itajub\'a, MG, 37500-000, Brazil 
}

\date{Received dd-mm-yy / Accepted dd-mm-yy}

\abstract
{Protoplanetary nebulae typically present non-spherical envelopes. The origin of such geometry is still controversial. There are indications that it may be carried over from an earlier phase of stellar evolution, such as the AGB phase. But how early in the star's evolution does the non-spherical envelope appear? }
{Li-rich giants show dusty circumstellar envelopes that can help answer that question. We study a sample of fourteen Li-rich giants using optical polarimetry in order to detect non-spherical envelopes around them. }
{We used the IAGPOL imaging polarimeter to obtain optical linear polarization measurements in ${\it V}$ band. Foreground polarization was estimated using the field stars in each CCD frame.}
{After foreground polarization was removed, seven objects presented low intrinsic polarization (0.19 $-$ 0.34)\% and two (\object{V859 Aql} and \object{GCSS 557}) showed high intrinsic polarization values (0.87 $-$ 1.16)\%. This intrinsic polarization suggests that Li-rich giants present a non-spherical distribution of circumstellar dust. The intrinsic polarization level is probably related to the viewing angle of the envelope, with higher levels indicating objects viewed closer to edge-on. The correlation of the observed polarization with optical color excess gives additional support to the circumstellar origin of the intrinsic polarization in Li-rich giants. The intrinsic polarization correlates even better with the IRAS  25 $ \mu $m far infrared emission. Analysis of spectral energy distributions for the sample show dust temperatures for the envelopes tend to be between 190 and 260 K. We suggest that dust scattering is indeed responsible for the optical intrinsic polarization in Li-rich giants.}
{Our findings indicate that non-spherical envelopes may appear as early as the red giant phase of stellar evolution. }

\keywords{polarization -- stars: circumstellar matter -- infrared -- stars: atmospheric parameters -- stars: abundances -- stars: late type giants -- stars: evolution -- element: Li }

\maketitle \markboth{Pereyra et al.: Polarimety of Li-rich giants }{}

\section{Introduction.}
Mass loss plays a central role in the late stages of stellar evolution. In particular, asymptotic giant branch (AGB) stars may be responsible for up to 60\% of the (all stars) interstellar dust input into the interstellar medium (Gehrz \cite{gehr89}). 

Our current understanding of stellar evolution places the AGB stars as precursors of protoplanetary nebulae (PPN). On the other hand, we know from direct imaging that PPN typically present non-spherically symmetric envelopes. While extrinsic (to the PPN) effects such as binarity have been said to cause the non-sphericities (for a review see Balick \& Frank \cite{bali02}), could it not be instead that the asymmetry is carried over from the earlier, AGB phase of stellar evolution? 

One piece of evidence showing that departures from spherical symmetry exist around AGB stars is that light from these objects may show some degree of linear polarization (Coyne \& Magalh\~aes \cite{coyn77,coyn79}; Magalh\~aes et al. \cite{maga86a,maga86b}; Kahane et al. \cite{kaha97}; Magalh\~aes \& Nordsieck \cite{maga00}). Further evidence of asphericity in the envelopes of late-type giant and supergiant star envelopes comes from the details of OH maser emission profiles (Collison \& Fix \cite{coll92}), KI resonant scattering (Plez \& Lambert \cite{plez94}), rings of SiO maser emission (Diamond et al. \cite{diam94}, Greenhil et al. \cite{gree95}), and OH radio images (Chapman et al. \cite{chap94}).

Aspherical symmetries, such as those found by Trammell et al. (1994) in post-AGB stars from spectropolarimetry, may then be understood naturally, since such symmetries are already present in earlier evolutionary stages. The observed and more obvious non-spherical symmetries in PPN  (Balick \& Franck \cite{bali02}) and the inferred assymetries in other evolved objects obtained from polarimetry (Johnson \& Jones \cite{john91}, Parthasarathy \& Jain \cite{part93}, Parthasarathy et al. \cite{part05}) are also consistent with the origin of the asymmetries early in the AGB phase.

Aside from the fact that polarimetry {\it per se} allows the study of otherwise unresolved objects, the question arises as to how early in stellar evolution the non-spherical symmetry appears. In this paper we use the fact that Li-rich red giants (RG) show dusty circumstellar envelopes to explore their environments and to help answer that question.

Over the past two decades around 40 red giants were found to have Li abundances that are 100 times larger than the mean values observed in red giants (Brown et al. \cite{brow89}, de la Reza \& Drake \cite{dela95}). This number indicates that about 2$\%$ of the Population I red giants show significantly larger lithium abundances than expected by dilution due to mixing by classical convection. In some of the giants, the lithium abundance reaches values that are similar to (and even larger than) the Pop I value (meteoritic, open clusters, etc.), around logN(Li) = 3.3. Castilho et al. (\cite{cast99a}) obtained very low Be abundances for two Li-rich giants (LRG) providing evidence that the original Li in these stars must have been almost completely destroyed and that the high Li abundances in the Li-rich red giants are due to Li production in these stars. 

It seems to be clear that LRG are quite normal stars, except for their high Li abundance and large infrared excess (Castilho et al. \cite{cast95}, de la Reza et al. \cite{reza96}, Castilho et al. \cite{cast00}). The similarities to normal giants, namely mass, chemical composition, temperature, and metallicity, combined with the far-infrared emission, indicate that LRG do not form a unique class of objects but are ordinary low-mass stars observed during a short phase of their evolution, when Li is created.

If indeed all low mass red giants go through a phase of Li production in the RGB (that could be cyclic), together with an increase in mass loss seen in the IRAS color diagram (Gregorio-Hetem et al. \cite{greg93}, de la Reza et al. \cite{reza96}), and since some of them reach a larger lithium abundance than the Pop I abundance, they could be an important source of Li enrichment in the Galaxy. Some information needed to quantify the LRG contribution for the interstellar medium Li enrichment remains unknown, such as: the maximum Li abundance reached for each star and its relation with stellar mass and/or metallicity, the duration of the Li production phase(s), the mass loss process, and rate, and  the simultaneity of the envelope ejection with the Li production. 

Up to now only one envelope of LRG has been studied in detail. The $\sim$ 3x4 arcmin envelope of \object{HD 65750} (Castilho et al. \cite{cast98}) has a butterfly geometry and the present mass loss rate of the star is not enough to form the observed envelope. Witt \& Rogers (\cite{witt91}) proposed that a past and more efficient mass-loss event about 32,000 years ago was responsible for the observed structure.

The aim of this work is to measure the polarization of LRG stars so as to study the spatial distribution of the circumstellar dust around these objects. Scattering of the stellar radiation by the circumstellar dust can produce polarization that is measurable in objects with non-resolved envelopes. The amount of polarization will depend upon the density and nature of the scatterers, as well as on the aspect angle and exact geometry of the envelope (Magalh\~aes \cite{maga92}).  If intrinsic polarization is detected, we can infer that an asymmetric spatial distribution of circumstellar dust is present in these objects.

Here we present the results of optical linear polarization measurements of fourteen LRG and two normal giants (NG). In Sect. 2 we describe the observations and data reduction along with the calculations of foreground and intrinsic polarizations. In Sect. 3 we show the stellar parameters of our sample and a general discussion is presented including correlations between the polarimetric data and optical and near infrared excess color. Correlation with IRAS colors also are explored and spectral energy distributions used to estimate the dust temperature associated to the circumstellar envelopes. The conclusions are drawn in Sect. 4. 

\begin{figure}
\resizebox{\hsize}{!}{\includegraphics{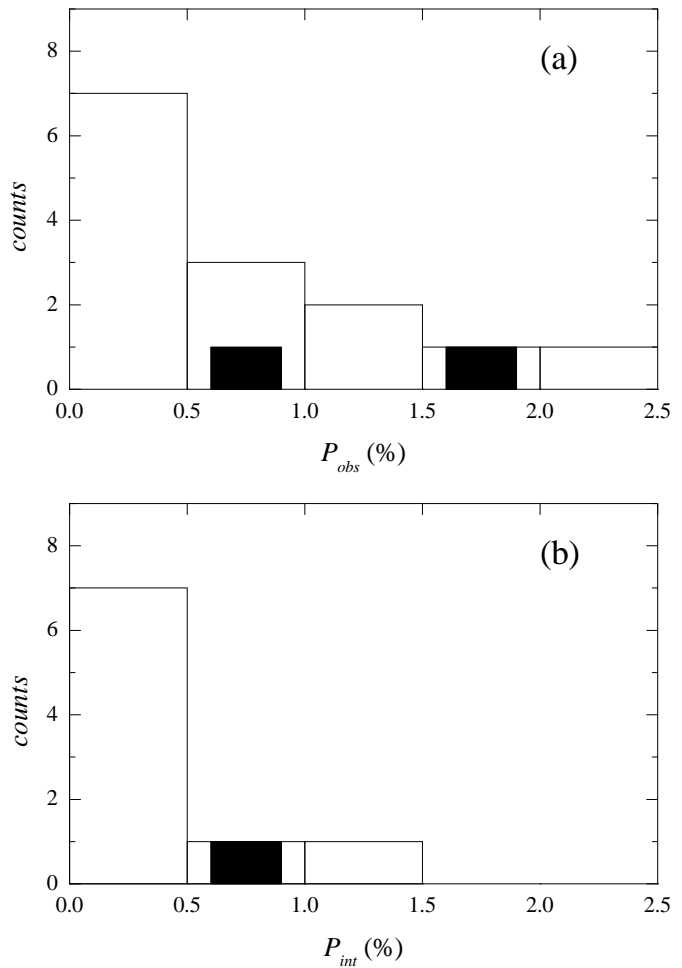}}
\caption{Histograms of (a) the observed and (b) intrinsic polarization of LRG. The NG in the sample are indicated in black.}
\label{histo}
\end{figure}


\begin{table*}
\caption{Log of observations.\label{log}}
\begin{tabular}{lccccccccc}
\hline \hline 
Object     & Log $ \varepsilon $(Li)$^{a}$     & {\it l} & {\it b} &  {\it V}$^{b}$  &  Wav. pos. &  IT &  Aper. & Date & ID \\
    & (dex) & &   & (mag) &  & (sec.) & (\arcsec)  & \\
    (1) & (2)  & (3) & (4) & (5) & (6) & (7) & (8) & (9) & (10) \\
\hline
\multicolumn{10}{c}{Li-rich giants}\\
\hline
\object{HD 90082}	&	0.20	(0.15)	&	286.27	&	-3.91	&	7.50	&	16	&	8	&	5.5	&	06/20/98	&	a	\\
\object{HD 95799}	&	3.10	(0.15)	&	289.19	&	1.17	&	8.14	&	16	&	12	&	5.5	&	06/20/98	&	b	\\
\object{HD 96195}	&	0.40	(0.20)	&	291.12	&	-2.57	&	7.94	&	16	&	13	&	6.7	&	06/20/98	&	c	\\
\object{HD 120602}	&	1.90	(0.15)	&	338.54	&	64.21	&	6.01	&	16	&	1	&	3.7	&	06/20/98	&	d	\\
\object{PDS 68}	&	2.00	(0.40)	&	315.72	&	19.10	&	12.80	&	4	&	300	&	4.3	&	06/20/98	&	e	\\
\object{HD 146850}	&	1.60	(0.15)	&	359.59	&	24.44	&	6.10	&	6$^{c}$	&	1	&	3.7	&	06/20/98	&	f	\\
\object{GCSS 557} (\object{V385 Sct})	&	0.70	(0.15)	&	17.10	&	-1.37	&	13.26	&	4	&	300	&	3.7	&	06/20/98	&	g	\\
\object{HD 176588}	&	1.60	(0.15)	&	30.08	&	-4.18	&	6.89	&	16	&	2	&	4.9	&	06/20/98	&	h	\\
\object{IRAS 19012-0747}	&	2.60	(0.15)	&	27.46	&	-6.28	&	11.17	&	16	&	60	&	3.1	&	06/20/98	&	i	\\
\object{IRAS 19038-0026}	&	0.60	(0.25)	&	34.32	&	-3.49	&	14.48	&	8	&	300	&	3.7	&	06/20/98	&	j	\\
\object{HD 178168}	&	0.90	(0.15)	&	32.54	&	-4.72	&	9.06	&	16	&	20	&	5.5	&	06/20/98	&	k	\\
\object{HD 112127}	&	2.80	(0.15)	&	0.96	&	89.34	&	6.87	&	16	&	1	&	4.9	&	06/21/98	&	l	\\
\object{V859 Aql} (\object{PDS 100})	&	2.50	(0.15)	&	42.29	&	-6.28	&	10.44	&	16	&	40	&	4.9	&	06/21/98	&	m	\\
\object{HD 203251}	&	1.40	(0.15)	&	35.57	&	-39.97	&	8.00	&	16	&	8	&	7.3	&	06/21/98	&	n	\\
\hline
\multicolumn{10}{c}{normal giants}\\
\hline
\object{HD 190664}	&	$<$0.00 (0.30)	&	37.92	&	-18.48	&	6.47	&	16	&	1	&	3.7	&	06/21/98	&	o	\\
\object{HD 124649}	&	$<$0.00 (0.30)	&	315.53	&	7.45	&	7.86	&	16	&	6	&	6.1	&	06/21/98	&	p	\\
\hline
\end{tabular}																			
\\
Notes: (a) log $ \varepsilon $(Li) were obtained from Castilho et al. \cite{cast00}, except for: \object{HD 120602} and \object{HD 112127} (from Brown et al. \cite{brow89}), \object{HD 146850} (from Castilho et al. \cite{cast95}), \object{V959 Aql} (from Reddy et al. \cite{redd02}), and \object{HD203251} (from Fekel \& Balachandran \cite{feke93}). The value for \object{PDS 68} is an estimation based on the equivalent width of the Li I (6707.8\AA) line (from Gregorio-Hetem et al. \cite{greg92}), and the values for \object{HD 190664} and \object{HD 124649} are higher limits (Castilho et al. \cite{cast99a}). The errors are indicated in parenthesis; (b) {\it V} were obtained from Castilho (\cite{cast99b}), except for: \object{HD 120602}, \object{HD 112127}, \object{HD 203251}, and \object{HD 190664} (from GCPC, Mermilliod et al. \cite{merm97}), \object{PDS 68} and \object{V859 Aql} (from Gregorio-Hetem et al. \cite{greg92}), \object{HD 124649} from CDS, and \object{IRAS 19038-0026} (from the GSC 2.2 catalogue, using magnitude in the visual phographic band as {\it V}); (c) 8 waveplate positions were initially observed but two of them were saturated.
\label{log}
\end{table*}

\begin{figure}
\resizebox{\hsize}{!}{\includegraphics{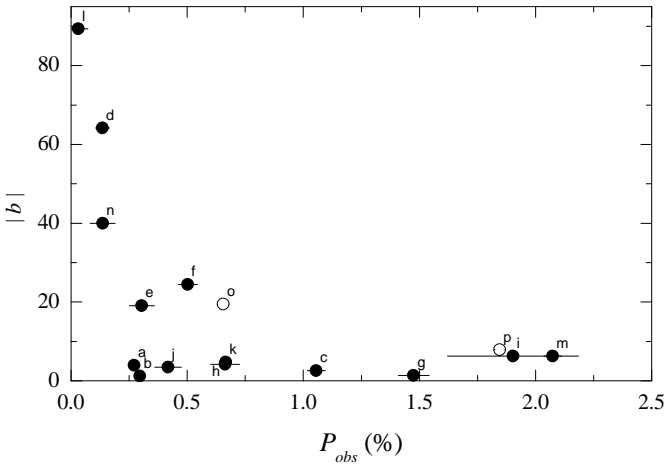}}
\caption{Correlation between observed polarization and galactic latitude. The black dots are the LRG, and the white dots are the NG. The letters indicate the ID for each object as it appears in column 10 of Table \ref{log}.}
\label{lat}
\end{figure}

\section{Observations and data reduction\label{obs}}

The observations were made using IAGPOL, the IAG imaging polarimeter (Magalh\~aes et al. \cite{maga96}), at the f/13.5 Cassegrain focus of the 0.6m IAGUSP Telescope at the Observat\'orio do Pico dos Dias (OPD), operated by the Laborat\'orio Nacional de Astrof\' {\i}sica (LNA), Brazil. When in linear polarization mode, the polarimeter incorporates a rotatable, achromatic half-wave retarder followed by a calcite Savart plate. This provides two images of each object in the field, separated by 1mm (corresponding to 25.5$\arcsec$ at the telescope's focal plane), with orthogonal polarizations. One polarization modulation cycle is covered for every 90$\degr$ rotation of the waveplate. The simultaneous imaging of the two beams allows observation under non photometric conditions, and at the same time the sky polarization is practically canceled. The polarimeter is highly efficient and is photon-noise-limited for point sources. Additional details about this polarimeter can be found in Magalh\~aes et al. (\cite{maga96}), Pereyra (\cite{pere00}) and Pereyra \& Magalh\~aes (\cite{pere02,pere04}).


\begin{table*}
[ht]
\caption{Optical linear polarization measurements.\label{data}}
\begin{tabular}{lcccccccc}
\hline \hline 
Object    &  {\it P}$_{obs}$  &  $ \theta $$_{obs}$ &  {\it P}$_{for}$ & $ \theta $$_{for}$ & {\it N}  &  {\it P}/$\sigma$$_{{\it P}}$ &  {\it P}$_{int}$  &$ \theta $$_{int}$ \\
    &  (\%) & (\degr) & (\%) & (\degr) &  &  & (\%) & (\degr) \\
    (1) & (2)  & (3) & (4) & (5) & (6) & (7) & (8) & (9) \\
\hline
\object{HD 90082}	&	0.272	(0.014)	&	109.43	&	0.649	(0.028)	&	116.00	&	9	&	5	&	0.389	(0.031)	&	30.57	\\
\object{HD 95799}	&	0.296	(0.021)	&	96.03	&	0.440	(0.014)	&	111.93	&	17	&	5	&	0.245	(0.025)	&	41.72	\\
\object{HD 96195}	&	1.056	(0.039)	&	90.73	&	1.127	(0.063)	&	97.99	&	10	&	5	&	0.285	(0.074)	&	42.17	\\
\object{HD 120602}	&	0.135	(0.030)	&	70.23	&					&		&		&		&					&		\\
\object{PDS 68}	&	0.305	(0.054)	&	56.53	&	0.104	(0.001)	&	105.38	&	32	&	30	&	0.335	(0.054)	&	47.59	\\
\object{HD 146850}	&	0.502	(0.041)	&	102.13	&					&		&		&		&					&		\\
\object{GCSS 557}	&	1.475	(0.066)	&	43.03	&	0.831	(0.001)	&	58.37	&	222	&	20	&	0.871	(0.066)	&	28.46	\\
\object{HD 176588}	&	0.663	(0.062)	&	66.23	&					&		&		&		&					&		\\
\object{IRAS 19012-0747}	&	1.903	(0.282)	&	14.33	&	1.786	(0.017)	&	16.60	&	26	&	10	&	0.187	(0.283)	&	169.77	\\
\object{IRAS 19038-0026}	&	0.418	(0.057)	&	65.23	&	0.351	(0.006)	&	88.38	&	97	&	5	&	0.309	(0.057)	&	37.56	\\
\object{HD 178168}	&	0.666	(0.022)	&	59.43	&	0.781	(0.018)	&	47.56	&	9	&	5	&	0.318	(0.028)	&	108.87	\\
\object{HD 112127}	&	0.032	(0.041)	&	65.03	&					&		&		&		&					&		\\
\object{V859 Aql}	&	2.074	(0.038)	&	80.63	&	0.989	(0.019)	&	89.10	&	38	&	5	&	1.164	(0.043)	&	73.47	\\
\object{HD 203251}	&	0.136	(0.054)	&	143.13	&					&		&		&		&					&		\\
\hline
\object{HD 190664}	&	0.655	(0.026)	&	86.03	&					&		&		&		&					&		\\
\object{HD 124649}	&	1.846	(0.021)	&	63.23	&	2.377	(0.076)	&	68.99	&	16	&	5	&	0.677	(0.079)	&	175.48	\\

\hline
\end{tabular}																			
\\
The polarization errors are indicated in parenthesis.
\end{table*}

\begin{figure*}

$\begin{array}{ccc}

\resizebox{6cm}{!}{\includegraphics{4270f03a.eps}} &
\resizebox{6cm}{!}{\includegraphics{4270f03b.eps}} &
\resizebox{6cm}{!}{\includegraphics{4270f03c.eps}} \\
\resizebox{6cm}{!}{\includegraphics{4270f03d.eps}} &
\resizebox{6cm}{!}{\includegraphics{4270f03e.eps}} &
\resizebox{6cm}{!}{\includegraphics{4270f03f.eps}} \\
\resizebox{6cm}{!}{\includegraphics{4270f03g.eps}} &
\resizebox{6cm}{!}{\includegraphics{4270f03h.eps}} &
\resizebox{6cm}{!}{\includegraphics{4270f03i.eps}} \\
\resizebox{6cm}{!}{\includegraphics{4270f03j.eps}} & & \\
\end{array}$
\hfill
\put (-335,-340) {
\parbox[b]{11cm}{
\caption{Polarization maps for each target where foreground polarization was calculated. Each vector represents one object included in the average for each case (see Table \ref{data}). The polarization scale is shown to the upper right of each map. The maps are centered on the coordinates of each target with its observed polarization also shown. The coordinates are 2000.}}}
\label{vecplot}
\end{figure*}

The data were collected on 20 and 21 June 1998. The measurements were made using a standard {\it V} filter with a 1024x1024 CCD that covers an area of $ \sim $ 10'$\times$10'. Typical sequences of four, eight, or sixteen positions separated by 22.5$\degr$ of the half-waveplate were used depending on the magnitude of the star. Typical integration times for position of the half-waveplate were a few seconds for the brighter stars and a few minutes for the dimmer stars. Our sample, along with a log of observations, is shown in Table \ref{log}. The total sample is composed of sixteen objects and includes fourteen LRG and two NG. Th LRG were obtained from de la Reza \& Drake (\cite{dela95}) and Castillho et al. (\cite{cast98}). The object names are indicated in column (1) with their Li abundances in column (2). The galactic coordinates are in columns (3) and (4). The visual magnitudes are in column (5) and were obtained (when available) from the literature. In particular, \object{IRAS 19038-0026}, with no previous {\it V} magnitude published, was identified as the \object{S300232037086} object in the GSC 2.2 catalogue, and the photographic bands for this object were used in this compilation. The number of waveplate positions used in each object is indicated in column (6) and the integration time used by waveplate position is in column (7).  The radius of the aperture, which minimized the polarization error, used is shown in column (8). The date of each observation is indicated in column (9) and identification for each object is shown in the last column.

For the data reduction process, we follow the procedure indicated in Pereyra \& Magalh\~aes (\cite{pere02}) using the PCCDPACK package (Pereyra \cite{pere00}). This is a set of specially developed IRAF scripts to study the polarization data in (eventually crowded) stellar fields. The linear observed polarization ({\it P}) and the polarization position angle ($ \theta $, measured from north to east) for the sample are shown in columns (2) and (3) of Table \ref{data}. Corrections of the polarization position angle to the equatorial system were obtained from polarized standard stars observed each night. Unpolarized standard stars were used to check the instrumental polarization, which was found to be smaller than 0.04\%, so no correction for instrumental polarization was applied to the data.

One star in our sample, LRG: \object{HD 146850}, is also present in the catalogue of stellar polarization by Heiles (\cite{heil00}). The Heiles values ({\it P} = 0.61$ \pm $0.04\%, $ \theta$ =  89.1$ \pm $1.6$ \degr $) are consistent with our {\it P} value, but $\theta$ presents a discrepancy that might indicate some variability. Our polarization measurements will be considered for the discussion in the next section.

Figure \ref{histo}a shows the histogram of observed polarization for the stars in our sample. Approximately half of the LRG stars have observed polarizations that are lower than 0.5\%, but in a few cases high polarizations ($>$1.5\%) are observed (\object{IRAS 19012-0747} and \object{V859 Aql}), and one of the normal giants has an enhanced polarization value ($ \sim $ 1.8\%, \object{HD 124649}).

\subsection{Foreground polarization\label{fore}}

The foreground polarization must be important if, for example, the line of sight is along the Galactic plane. Figure \ref{lat} shows the observed polarization for our sample as a function of the Galactic latitude modulus. Clearly, for the objects with $\mid${\it b} $\mid$ $<$ 20$\degr$, the polarization covers a wide range. For higher latitudes, the observed polarization presents lower values ($<$ 0.5\%). Thus, for half of our sample the contribution of foreground polarization is important and must be subtracted. It is interesting to note that the objects with the lowest observed polarizations (LRG: \object{HD 112127}, \object{HD 120602}, and \object{HD 203251}) are at higher latitudes. We can conclude that the observed polarizations of these objects represent the very low or null foreground polarization in these directions.



\begin{table*} 
\caption{Sample stellar parameters.\label{para}}
\setlength\tabcolsep{3pt}
\begin{tabular}{lccccccccccr@{ }lc}
\hline \hline 
Object     & T$_{eff}$$^{a}$    & Log {\it g}$^{a}$ & [Fe/H]$^{a}$ & {\it U} -{\it B}$^{b}$ & {\it B} -{\it V}$^{b}$ &{\it V} -{\it R}$^{b}$ &{\it R} -{\it I}$^{b}$ & ({\it B} -{\it V})$_{0}$$^{c}$ & {\it E}({\it B} -{\it V})  &  Plx$^{d}$  & \multicolumn{2}{c}{Distance} & {\it E}({\it B} -{\it V})$_{Bond}$ \\
    & & & & & &   &  & & & (mas) & \multicolumn{2}{c}{(pc)} & \\
    (1) & (2)  & (3) & (4) & (5) & (6) & (7) & (8) & (9) & (10) & (11) & \multicolumn{2}{c}{(12)} & (13)\\
\hline
\object{HD 90082}	&	3600	&	0.0	&	0.0	&	1.95	&	1.67	&	1.03	&	1.15	&	1.60	&	0.07	&	7.20	(4.80)	&	139	&	${+	278	\atop -	56	}$	&	0.03	\\
\object{HD 95799}	&	4900	&	2.5	&	0.0	&	0.79	&	1.00	&	0.53	&	0.51	&	0.95	&	0.05	&	20.50	(11.50)	&	49	&	${+	62	\atop -	18	}$	&	0.01	\\
\object{HD 96195}	&	3600	&	-0.5	&	0.0	&	1.61	&	2.28	&	1.42	&	1.56	&	1.74	&	0.54	&	1.12	(0.74)	&	893	&	${+	1739	\atop -	355	}$	&	0.18	\\
\object{HD 120602}	&	5000	&	3.0	&	-0.1	&	0.59	&	0.90	&	0.68	&	--	&	0.90	&	0.00	&	8.09	(0.81)	&	124	&	${+	14	\atop -	11	}$	&	0.00	\\
\object{PDS 68}	&	4450	&	2.6	&	0.0	&	1.90	&	1.58	&	0.88	&	0.79	&	1.14	&	0.44	&	--		--		&	--	&						&		\\
\object{HD 146850}	&	4000	&	1.5	&	-0.3	&	1.70	&	1.48	&	0.91	&	0.79	&	1.36	&	0.12	&	3.77	(0.85)	&	265	&	${+	77	\atop -	49	}$	&	0.04	\\
\object{GCSS 557}	&	3400	&	0.0	&	0.0	&	--	&	4.68	&	3.13	&	2.35	&	1.57	&	3.11	&	--		--		&	3113$^{e}$	&						&	0.56	\\
\object{HD 176588}	&	4000	&	1.5	&	0.0	&	2.10	&	1.65	&	0.97	&	0.88	&	1.38	&	0.27	&	4.25	(0.91)	&	235	&	${+	64	\atop -	41	}$	&	0.05	\\
\scriptsize{\object{IRAS 19012-0747}}	&	3800	&	1.5	&	0.0	&	2.30	&	1.83	&	1.02	&	0.93	&	1.39	&	0.44	&	--		--		&	2330$^{e}$	&						&	0.24	\\
\scriptsize{\object{IRAS 19038-0026}}	&	3600	&	1.0	&	0.0	&	--	&	1.20	&	--	&	--	&	1.35	&	neg.	&	--		--		&	--	&						&		\\
\object{HD 178168}	&	4000	&	1.0	&	0.0	&	2.17	&	2.03	&	1.09	&	0.97	&	1.43	&	0.60	&	17.70	(16.90)	&	56	&	${+	1194	\atop -	28	}$	&	0.01	\\
\object{HD 112127}	&	4340	&	2.1	&	0.3	&	1.43	&	1.26	&	0.90	&	0.58	&	1.26	&	0.00	&	8.09	(0.85)	&	124	&	${+	15	\atop -	12	}$	&	0.00	\\
\object{V859 Aql}	&	4500	&	2.5	&	0.0	&	1.44	&	1.31	&	0.82	&	0.73	&	1.12	&	0.19	&	--		--		&	--	&						&		\\
\object{HD 203251}	&	4500	&	3.0	&	-0.3	&	1.10	&	1.22	&	--	&	--	&	1.08	&	0.14	&	-0.20	(9.10)	&	plx	&	$<$ 0					&		\\
\hline
\object{HD 190664}	&	4650	&	2.7	&	0.0	&	1.00	&	1.16	&	0.86	&	--	&	1.06	&	0.10	&	9.57	(0.96)	&	104	&	${+	12	\atop -	10	}$	&	0.02	\\
\object{HD 124649}	&	3750	&	1.5	&	0.0	&	--	&	1.70	&	--	&	--	&	1.39	&	0.31	&	1.98	(0.97)	&	505	&	${+	485	\atop -	166	}$	&	0.09	\\

\hline
\end{tabular}
													
Notes: (a) obtained from Castilho et al. (\cite{cast00}), except for: \object{HD 120602} and \object{HD 112127} (from Brown et al. \cite{brow89}), \object{HD 146850} (from Castilho et al \cite{cast99a}), \object{V859 Aql} (from Reddy et al. \cite{redd02}), \object{HD 203251} (from Fekel \& Balachandran \cite{feke93}), and \object{PDS68}, \object{HD 190664}, and \object{HD 124649} (see text); (b) obtained from Castilho (\cite{cast99b}), except for: \object{HD 120602}, \object{HD 112127}, \object{HD 203251}, and \object{HD 190664} (from GCPC, Mermilliod et al. \cite{merm97}), \object{PDS 68} and \object{V859 Aql} (from Gregorio-Hetem et al. \cite{greg92}), \object{HD 124649} from CDS. The {\it B} -{\it V} color index for \object{GCSS 557} and \object{IRAS 19038-0026} were obtained from the GSC 2.2 catalogue, using magnitudes in {\it J} and {\it V} photographic bands as {\it B} and {\it V}, respectively; (c) obtained from Bessell et al. (\cite{bess98}) with interpolation (or extrapolation) of T$_{eff}$, log {\it g} and [Fe/H] (columns (2), (3), and (4)); (d) obtained from the Hipparcos catalogue, except for: \object{HD 90082}, \object{HD 95799}, \object{HD 178168}, and \object{HD 203251} (from the Tycho catalogue). The parallax errors are indicated in parenthesis; (e) from Castilho et al. (\cite{cast00}). 

\label{optical}
\end{table*}


The foreground polarization for the objects in our sample was estimated with the field stars in each CCD frame using PCCDPACK. This package provides the polarization for each object present in a typical CCD frame, as well as the average Stokes parameters ({\it Q} and {\it U}) weighted by the errors for all the objects considered. With these parameters the average foreground polarization can be calculated in a field of view that is very close to the target and just limited by the CCD size. We used a polarization signal-to-noise ratio ({\it P}/$\sigma$$_{{\it P}}$) of 5 or larger in order to select objects to be included in the average (see Fig. \ref{vecplot}). We take this average as an estimate of the foreground polarization value. These are indicated in columns (4) and (5) of Table \ref{data}, along with the number of objects and the minimum threshold {\it P}/$\sigma$$_{{\it P}}$ used in each case in columns (6) and (7). For bright objects, which required short integration times, and/or for those with higher latitudes, it was usual to find no field stars on the CCD frame. In these cases, it was not possible to obtain the foreground polarization, so those values are left blank in Table \ref{data}. In three of these cases (LRG: \object{HD 112127}, \object{HD 120602}, and \object{HD 203252}), the observed polarization has a {\it P}/$\sigma$$_{{\it P}}$ lower than 5 and these objects can be considered unpolarized. These objects are those located at higher latitudes as mentioned above.

\subsection{Intrinsic polarization\label{intr}}

For those objects with an estimated foreground polarization (Table \ref{data}), it was possible to obtain the intrinsic polarization. As the foreground polarization is an additive component included in the observed polarization, the intrinsic Stokes parameters ({\it Q} =  {\it P}$ \cos $(2$ \theta $) and {\it U} = {\it P}$ \sin $(2$ \theta $)) are as follows:

\vspace{0.5cm} 

{\it Q}$_{int}$   =   {\it Q}$_{obs}$ - {\it Q}$_{for}$     

{\it U}$_{int}$   =   {\it U}$_{obs}$ - {\it U}$_{for}$.   

\vspace{0.5cm} 

The intrinsic polarization ({\it P}$_{int}$) and its polarization angle ($ \theta $$_{int}$) are obtained from:

\vspace{0.5cm} 

{\it P}$_{int}$ =  ({\it Q}${2 \atop int}$ + {\it U}${2 \atop int}$)$^{1/2}$

$ \theta $$_{int}$ = $ \frac{1}{2}$$ \arctan $({\it U}$_{int}$/{\it Q}$_{int}$).

\vspace{0.5cm} 

The intrinsic polarization estimated for ten objects (9 LRG and 1 NG) are indicated in columns (8) and (9) of Table \ref{data} and shown in Fig. \ref{histo}b. In general, the intrinsic polarization levels are significant. Seven LRG (five of them with {\it P}/$\sigma$$_{{\it P}}$ $>$ 5) have intrinsic polarizations between (0.19-0.34)\%, but in some cases high values ($>$ 0.5\%) are obtained (LRG: \object{GCSS 557} and \object{V859 Aql}; and NG: \object{HD124649}). As an extreme case, one of the objects with high observed polarization (\object{IRAS 19012-0747}, 1.90\%) presents negligible intrinsic polarization within 5$\sigma$ (0.19\%).

A quantitative analysis of intrinsic polarization is beyond the scope of this paper; nevertheless, the polarization
models of circumstellar dust shells from Johnson and Jones (\cite{john91}) are useful for investigating the geometry of the shells in our sample. In that work, the shells around evolved stars were modeled as ellipsoids of revolution and the results pointed out the importance of taking into account the inclination of the ellipsoid to the plane of the sky ({\it i'}). For an ellipsoid of any given axial ratio, the inclination would decrease the polarization by $\sim$ cos$^{2}$({\it i'}) (or $\sim$ sin$^{2}$({\it i}), where {\it i} is the inclination to the line of sight). Therefore, a tilted ellipsoid would need a higher axial ratio to produce the same amount of polarization as a nontilted ellipsoid would. Comparing the red giant edge-on models in Johnson and Jones (\cite{john91}), polarization of  $\sim$ 0.3\%  was found for slightly prolate optically thin shells (as in \object{R Vir}) and $\sim$ 2.3\% for prolate optically thick shells (as in \object{RU Vir}). This last case is more consistent with the full range of intrinsic polarization in our sample [0.19$-$1.16]\%. Assuming the dependence with sin$^{2}$({\it i}), our sample is best represented by pole-on objects ({\it i} $<$ 30$\degr$), except for \object{GCSS 577} and \object{V859 Aql}. These two objects have a high chance of being closer to edge-on (probably,  {\it i} $>$ 45$\degr$). A more detailed modeling using Monte Carlo methods developed by our group (Carciofi et al. \cite{carc04}) is planned.

\section{Discussion}

\subsection{{\it E}({\it B} -{\it V}) and distances}

Correlations between the observed polarizations and the color excess can also help to investigate if an intrinsic color excess is associated with intrinsic polarization. This can be especially important for objects with undetermined intrinsic polarization. For this purpose, we computed the color excess, {\it E}({\it B} -{\it V}), for the objects in our sample, using the information in the literature when available (see Table \ref{para}). 

The effective temperatures, surface gravities, and metallicities are shown in columns (2), (3), and (4) of Table \ref{para}. Three objects (\object{PDS 68}, \object{HD 190664}, and \object{HD 124649}) have unknown stellar parameters to date and the listed values were calculated following Castilho et al. (\cite{cast00}) using the near infrared colors (see Sect. \ref{nirsec}). 

\begin{figure}
\resizebox{\hsize}{!}{\includegraphics{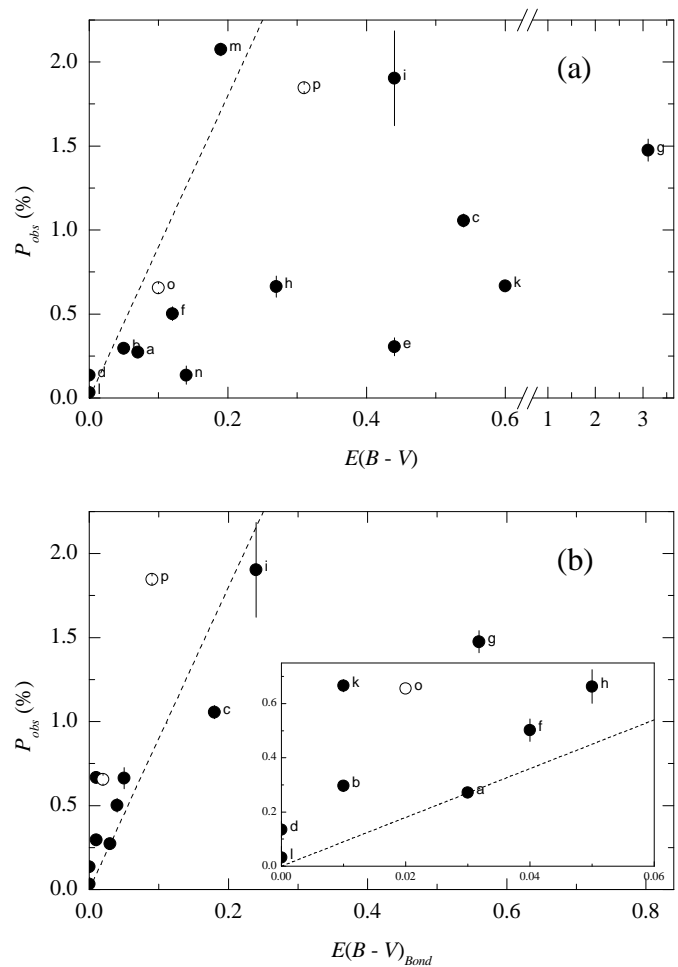}}
\caption{Correlation between observed polarizations and color excess (a) using intrinsic colors, and (b) using the Bond extinction law. The black dots are the LGR, and the white dots are the NG. The dashed line represents the upper limit for optimum alignment for polarizing dust grains in the diffuse ISM ({\it P}$_{V}$/{\it E}({\it B} -{\it V}) = 9\% mag$^{-1}$, Serkowski et al. \cite{serk75}). The box inside (b) figure is a zoom to show more detail. The letters indicate the ID for each object as it appears in column 10 of Table \ref{log}.}
\label{exce}
\end{figure}

\begin{figure}
\resizebox{\hsize}{!}{\includegraphics{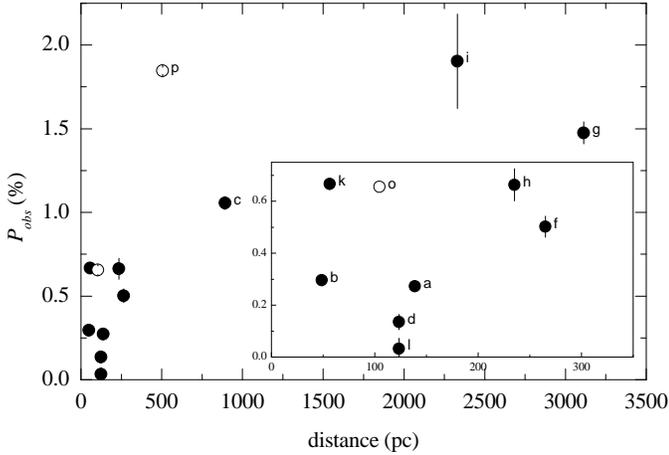}}
\caption{Correlation between observed polarizations and distance. The black dots are the LGR, and the white dots are the NG. The box inside the figure is a zoom to show more detail. The letters indicate the ID for each object as it appears in column 10 of Table \ref{log}.}
\label{dist}
\end{figure}

The color indexes {\it U} -{\it B}, {\it B} -{\it V}, {\it V} -{\it R}, and {\it R} -{\it I}, when available, are shown in columns (5) to (9) of Table \ref{para}, along with the intrinsic color index ({\it B} -{\it V})$_{0}$. In particular, \object{GCSS 557}, with no previously published {\it B} -{\it V} index, was identified as \object{S300112223323} in the GSC 2.2 catalogue and its {\it B} -{\it V} color index compiled here was constructed using the available photographic bands ({\it J} and {\it V}, respectively). As a comparison, the visual magnitude {\it V}=13.26 from Castilho (\cite{cast99b}) agrees with the {\it V}=13.50 photographic band magnitude (from GSC 2.2). An important discrepancy is found between its color ({\it V} -{\it R} = 3.13) from Castilho (\cite{cast99b}) and the ({\it V} -{\it F} = 0.64) index using the red photographic band ({\it F}) from GSC 2.2. Then, the {\it B} -{\it V} index used here for this object must be taken with care. Finally, the calculated color excess, {\it E}({\it B} -{\it V}), for our sample is indicated in column (10) of Table \ref{para}. Negative values were not considered, as in case of \object{IRAS 19038-0026}.


\begin{table*} 
\caption{Infrared data.}
\setlength\tabcolsep{5pt}
\begin{tabular}{llccccccccl}
\hline \hline 
Object     & 2MASS ID   & {\it J}  &  {\it H}  & {\it K}  & Flag$^{a}$ & ({\it J}-{\it H})$_{0}$$^{b}$ & ({\it H}-{\it K})$_{0}$$^{b}$ & {\it E}({\it J}-{\it H}) & {\it E}({\it H}-{\it K}) & IRAS ID   \\
    &   & (mag)  & (mag) & (mag) & & & & \\
    (1) & (2)  & (3) & (4) & (5) & (6) & (7) & (8) & (9) & (10) & (11)\\
\hline
\object{HD 90082}	&	19062377-0021222 	&	5.43	&	4.48	&	3.98	&	ADD	&	0.94	&	0.20	&	0.02	&	0.30	&	10204-6135	\\
\object{HD 95799}	&	11022509-5845373 	&	6.21	&	5.71	&	5.55	&	AAA	&	0.54	&	0.04	&	-0.04	&	0.12	&	none	\\
\object{HD 96195}	&	11042916-6257439 	&	3.48	&	2.43	&	2.07	&	DDD	&	0.91	&	0.22	&	0.15	&	0.14	&	11024-6241	\\
\object{HD 120602}	&	13502469+0529498 	&	4.74	&	4.22	&	4.10	&	DDD	&	0.51	&	0.03	&	0.01	&	0.09	&	13479+0544	\\
\object{PDS 68}	&	13565727-4208098 	&	9.90	&	9.16	&	8.89	&	AAA	&	0.68	&	0.07	&	0.06	&	0.20	&	13539-4153	\\
\object{HD 146850}	&	16190043-1452220 	&	3.27	&	2.46	&	2.27	&	DDD	&	0.87	&	0.11	&	-0.06	&	0.08	&	16161-1445	\\
\object{GCSS 557}	&	18265916-1441487 	&	4.86	&	3.34	&	2.71	&	ECD	&	1.03	&	0.23	&	0.50	&	0.40	&	18241-1443	\\
\object{HD 176588}	&	19010951-0426201 	&	4.10	&	3.19	&	2.96	&	DDD	&	0.84	&	0.11	&	0.07	&	0.12	&	18585-0430	\\
\scriptsize{\object{IRAS 19012-0747}}	&	19035796-0742345 	&	6.73	&	5.63	&	5.16	&	AAA	&	0.92	&	0.14	&	0.18	&	0.33	&	19012-0747	\\
\scriptsize{\object{IRAS 19038-0026}}	&	19062377-0021222 	&	5.43	&	4.48	&	3.98	&	ADD	&	0.98	&	0.17	&	-0.02	&	0.33	&	19038-0026	\\
\object{HD 178168}	&	10220747-6151005 	&	4.00	&	3.03	&	2.79	&	DDD	&	0.82	&	0.12	&	0.15	&	0.12	&	19049-0234	\\
\object{HD 112127}	&	12535573+2646479 	&	5.00	&	4.34	&	4.12	&	DCE	&	0.70	&	0.09	&	-0.04	&	0.14	&	12514+2703	\\
\object{V859 Aql}	&	19310123+0523533 	&	7.57	&	6.82	&	6.60	&	AAA	&	0.66	&	0.06	&	0.09	&	0.15	&	19285+0517	\\
\object{HD 203251}	&	21212967-1509216 	&	5.83	&	5.18	&	5.02	&	AAA	&	0.67	&	0.06	&	-0.02	&	0.10	&	21187-1522	\\
\hline
\object{HD 190664}	&	20061222-0404414 	&	4.51	&	3.95	&	3.85	&	ECC	&	0.62	&	0.06	&	-0.06	&	0.03	&	20035-0413	\\
\object{HD 124649}	&	14162210-5319365 	&	4.84	&	3.78	&	3.65	&	DDD	&	0.92	&	0.14	&	0.14	&	-0.01	&	14130-5305	\\

\hline
\end{tabular}																			
\\
Notes: (a) 2MASS {\it JHK} photometric quality flag (from "A": very good, to "E": very poor); (b) from Bessell et al. (\cite{bess98}).
\label{nir}
\end{table*}

Figure \ref{exce}a shows the correlation between the observed polarization and the color excess, {\it E}({\it B} -{\it V}). The upper limit for optimum alignment of polarizing dust grains in the diffuse ISM is also shown ({\it P}$_{V}$/{\it E}({\it B} -{\it V}) = 9\% mag$^{-1}$, Serkowski et al. \cite{serk75}). All the objects, except \object{V859 Aql} and \object{HD 120602}, are under this limit, so we might conclude that the observed polarization has an interstellar origin. Nevertheless, the non-negligible polarization found in Sect. \ref{intr} tell us that a fraction of the excess color must originate in the circumstellar material in these stars. \object{HD 120602} has an {\it E}({\it B} -{\it V}) = 0, and its position in Fig. \ref{exce}a probably reflects the very low or null foreground polarization in this direction as we noted in Sect. \ref{fore}. In contrast, the case of \object{V859 Aql} could indicate a very important contribution of the circumstellar envelope to the excess color that is observed. Both \object{V859 Aql} and \object{GCSS 557} are the two LRG with the highest computed intrinsic polarization, and they have a high chance of presenting an asymmetric spatial distribution of circumstellar dust (envelope) probably viewed edge-on.

An alternative method to obtain the color excess is to apply the Bond (\cite{bond80}) extinction law, but a distance estimate is needed. Distances were derived from parallaxes given in the Hipparcos and Tycho catalogues when available. The parallaxes and distances are indicated in columns (11) and (12) of Table \ref{para}, respectively, but negatives values were not considered (\object{HD 203251}). Objects with parallax errors that produce distance errors that are higher than 100 pc must be  considered with care (\object{HD 90082}, \object{HD 96195}, \object{HD 178168}, and \object{HD 124649}). For \object{GCSS 557} and \object{IRAS 19012-0747}, with unknown parallaxes, the distances were obtained from Castilho et al. (\cite{cast00}), who use a color-magnitude diagram for Hipparcos field stars (Perryman et al. \cite{perr95}). We must note here that three objects (\object{HD 90082}, \object{HD 95799}, and \object{HD 178168}) with distances from Tycho catalogue show significant discrepancies with the (apparently overestimated) values found by Castilho et al. (\cite{cast00}) for these stars. In the following, our tabulated distances to these 3 stars will be used.

Figure \ref{dist} plots the observed polarization and distance. Seven LRG and one NG are located within the first 300 pc with observed polarizations lower than 0.75\%. It is interesting to note that \object{GCSS 557} with its high observed polarization (and also intrinsic polarizations) is located at an extreme distance. This is consistent with the high level of foreground polarization calculated in this direction. The important foreground contribution in the line of sight to \object{IRAS 19012-0747} (see Sect. \ref{intr}) is also consistent with the larger assumed distance (2330 pc) for this object and with the fact that it is located at low galactic latitude.

With distance information we can obtain the color excess using the Bond (\cite{bond80}) extinction law, {\it E}({\it B} -{\it V})$_{Bond}$. For objects with {\it b} $ > $ 60\degr, {\it E}({\it B} -{\it V})$_{Bond}$ = 0; for {\it b} $ < $ -60\degr, {\it E}({\it B} -{\it V})$_{Bond}$ = 0.03; and for $ \mid ${\it b}$ \mid $ $<$ 60\degr, {\it E}({\it B} -{\it V})$_{Bond}$ = 0.03csc{\it b}[1-exp(-0.008{\it r}sin{\it b})] where {\it r} is the distance in parsecs and {\it b} is the Galactic latitude (in absolute value). This is shown in column (13) of Table \ref{para}. For \object{GCSS 557} and \object{IRAS 19012-0747}, the results are consistent with Castilho et al. (\cite{cast00}). If we consider that {\it E}({\it B} -{\it V}) includes the contribution of the foreground ISM along with an intrinsic color excess (such as in Fig. \ref{exce}a), and if {\it E}({\it B} -{\it V})$_{Bond}$ represents just the contribution of foreground ISM to a given line of sight, we would expect that an intrinsic color excess could explain the intrinsic polarization observed. Fig. \ref{exce}b shows the correlation between the observed polarization and the color excess using the Bond law. We see that a large part of the LRG present an excess of polarization when we compare with the upper limit of polarization with an ISM origin (Serkowski et al. \cite{serk75}).

\begin{figure}
\resizebox{\hsize}{!}{\includegraphics{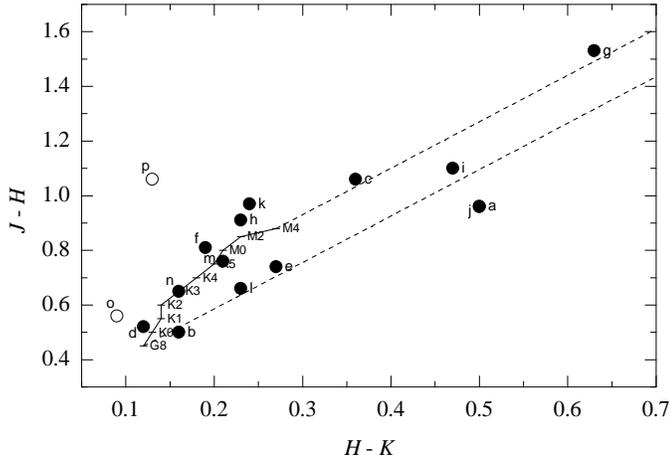}}
\caption{The ({\it J} - {\it H}) x ({\it H} - {\it K}) color-color diagram. The locus of the giant branch is shown by the solid line with spectral types indicated from Koornneef (\cite{koor83}) and the dashed lines follow the reddening vectors taken from Rieke \& Lebofsky (\cite{riek85}). The black dots are the LRG, and the white dots are the NG. The letters indicate the ID for each object as it appears in column 10 of Table \ref{log}.}
\label{ccdiag}
\end{figure}

Just three LRG with non-negligible intrinsic polarizations (\object{HD 96195}, \object{GCSS 557}, and \object{IRAS 19012-0747}) are under the  Serkowski limit, but in two cases (\object{GCSS 557} and \object{IRAS 19012-0747}) the inferred distance may have been overestimated as  mentioned earlier. If it is the case, {\it E}({\it B} -{\it V})$_{Bond}$ for these objects probably has a lower level and therefore these points are located more to the left. The distance error for \object{HD96195} obtained from its parallax error is bigger than 100 pc and also must be taken with care. 

Seven LRG and two NG have an observed polarization above the Serkowski limit, and in four of them (LRG: \object{HD 90082}, \object{HD 95799}, and \object{HD 178168}; and NG: \object{HD 124649}) we detected non-negligible intrinsic polarization. The two LRG with the lowest observed polarizations (\object{HD 120602} and \object{HD 112127}) are also slightly above this limit, but in both cases {\it E}({\it B} -{\it V})$_{Bond}$ = 0 because $ \mid ${\it b}$ \mid $ $ > $ 60\degr. This is consistent with our assumption of unpolarized objects given in Sect. \ref{obs}.

\begin{figure}
\resizebox{\hsize}{!}{\includegraphics{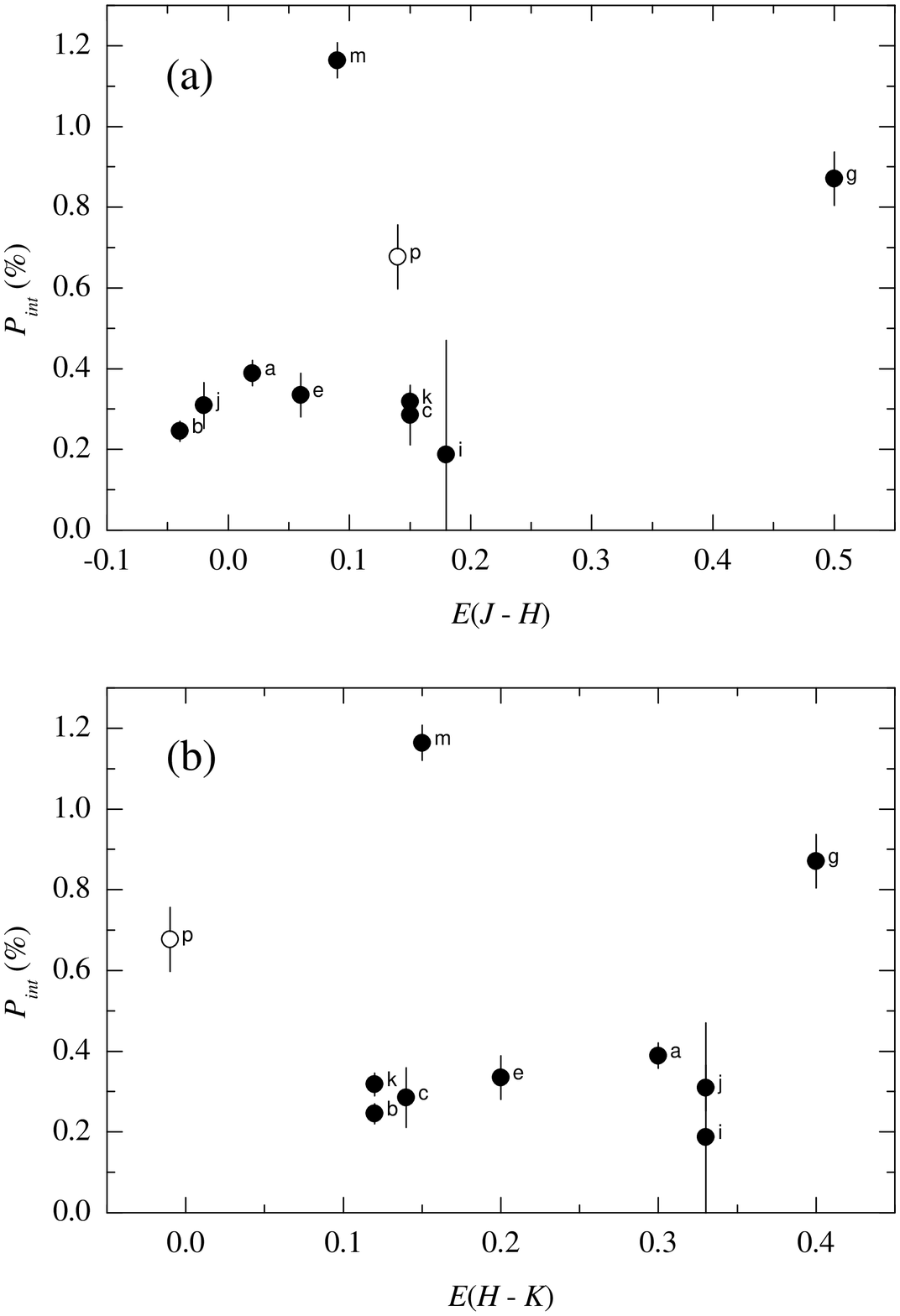}}
\caption{(a) Intrinsic polarization versus excess color {\it E}({\it J} - {\it H}); (b) Intrinsic polarization versus excess color {\it E}({\it H} - {\it K}). The black dots are the LRG, and the white dots are the NG. The letters indicate the ID for each object as appeared in column 10 of Table \ref{log}.}
\label{pnir}
\end{figure}

\begin{figure}
\resizebox{\hsize}{!}{\includegraphics{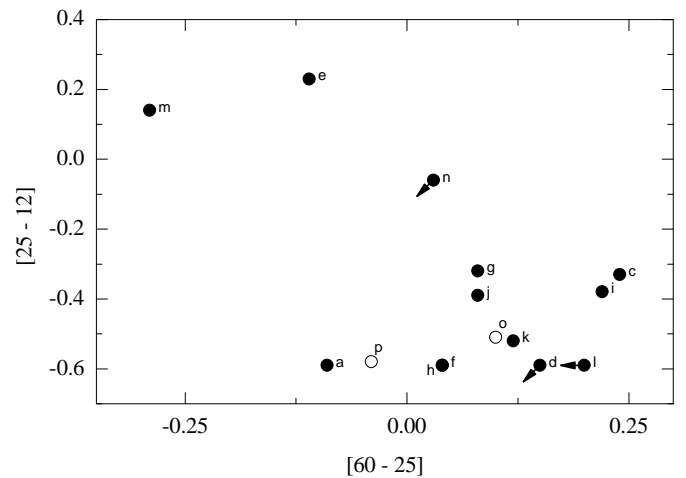}}
\caption{The IRAS color diagram, [25 - 12] versus [60 - 25]. The black dots are the LRG, and the white dots are the NG. The letters indicate the ID for each object as it appears in column 10 of Table \ref{log}.}
\label{iras}
\end{figure}

\begin{figure}
\resizebox{\hsize}{!}{\includegraphics{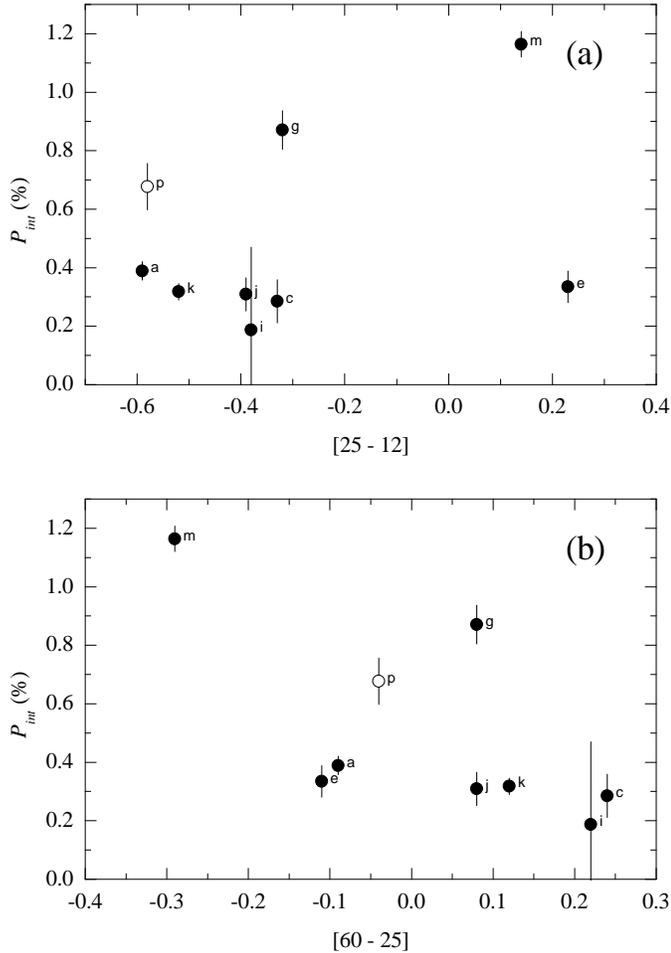}}
\caption{(a) Intrinsic polarization versus [25 - 12] IRAS color; (b) Intrinsic polarization versus [60 - 25] IRAS color. The black dots are the LRG, and the white dots are the NG. The letters indicate the ID for each object as it appears in column 10 of Table \ref{log}.}
\label{pir}
\end{figure}

It is interesting to note that three objects (LRG: \object{HD 146850}, \object{HD 176588}; NG: \object{HD 190604}) with significant observed polarization and without estimated foreground polarization  also appear above the Serkowski limit in Fig. \ref{exce}b. Thus, the correlation {\it P$_{obs}$} vs. {\it E}({\it B} -{\it V})$_{Bond}$ also gives information on a possible intrinsic color excess when an estimate of intrinsic polarization is not possible.

In general, analysis of the correlations of {\it P$_{obs}$} with the Bond law supports the idea that the non-negligible intrinsic polarization observed in LRG (and also in NG) has an origin in the circumstellar material.

\subsection{Correlations with infrared colors}

In a simple scenario, circumstellar dust in LRG must absorb light from the central object and radiate thermally in the infrared. Then, dust scattering will produce polarization, and a correlation between the optical intrinsic polarization and the excess (near or far) infrared emission is expected. Additionally, these correlations can give us information about the type of scatterer in the non-spherically symmetric envelope.

\subsubsection{Near infrared colors\label{nirsec}}

We used the 2MASS All-Sky Catalog of Point Sources (Cutri et al. \cite{cutr03}) to identify the NIR colors for the objects in our sample. The results are indicated in Table \ref{nir}. In columns (1) and (2), we show the named object and its NIR counterpart identified by the 2MASS ID, respectively. The {\it J}, {\it H}, and {\it K} colors are shown in columns (3), (4), and (5). Column (6) indicates the 2MASS {\it JHK} photometric quality flag for each measurement (from "A": very good, to "E": very poor). 

In Fig. \ref{ccdiag} we present the {\it J} - {\it K} versus {\it H} - {\it K} diagram for our sample, together with the positions of the giant branch (Koornneef \cite{koor83}) and the directions of the reddening vectors (Rieke \& Lebofsky \cite{riek85}) between G8III and M4III types. The majority of the objects have the colors of reddened photospheres, but two of LRG stars (\object{HD 90082} and \object{IRAS 19038-0026}) appear to the right of the reddening line for G8 giants, showing a slight excess at 2.2 $ \mu $m. With the NIR intrinsic colors obtained from Bessell et al. (\cite{bess98}) (cols. (7) and (8) in Table \ref{nir}), we calculated the NIR excess colors {\it E}({\it J} - {\it H}) and {\it E}({\it H} - {\it K}) (cols. (9) and (10) in Table \ref{nir}) for our sample using the stellar parameters from columns (2), (3), and (4) of Table \ref{para}.

The correlations between the intrinsic polarization and the NIR excess color are shown in Fig. \ref{pnir}. In order to serve as an objective criterion to quantify the correlation between {\it P}$_{{\it int}}$ and the NIR excess color, we choose the Spearman rank-order correlation coefficient ({\it r}$_{s}$), to which a statistical significance can be attached. The calculated value of {\it r}$_{s}$ for the {\it P}$_{{\it int}}$ versus {\it E}({\it J} - {\it H}) relation (Fig. \ref{pnir}a) is 0.10 with a 78\% probability, where a low probability means high significance.  For the {\it P}$_{{\it int}}$ versus {\it E}({\it H} - {\it K}) relation (Fig. \ref{pnir}b), {\it r}$_{s}$ is 0.07 with an 85\% probability. With this information, we can conclude that there is no statistically significant correlation between {\it P}$_{{\it int}}$ and NIR excess color. \object{V859 Aql} and \object{GCSS 557} with the highest intrinsic polarizations present low and high NIR excess color values, respectively; while \object{HD 90082} and \object{IRAS 19038-0026} with an excess in 2.2 $ \mu $m (Fig.  \ref{ccdiag}) have lower intrinsic polarization values. Using the NIR results, we can conclude that NIR emission either (a) is not responsible for the optical {\it P}$_{{\it int}}$ in LRG or (b) is responsible for the optical {\it P}$_{{\it int}}$ in LRG and the majority of LRG in our sample are pole-on objects except \object{V859 Aql} and \object{GCSS 557}.

\begin{figure*}
\resizebox{\hsize}{!}{\includegraphics{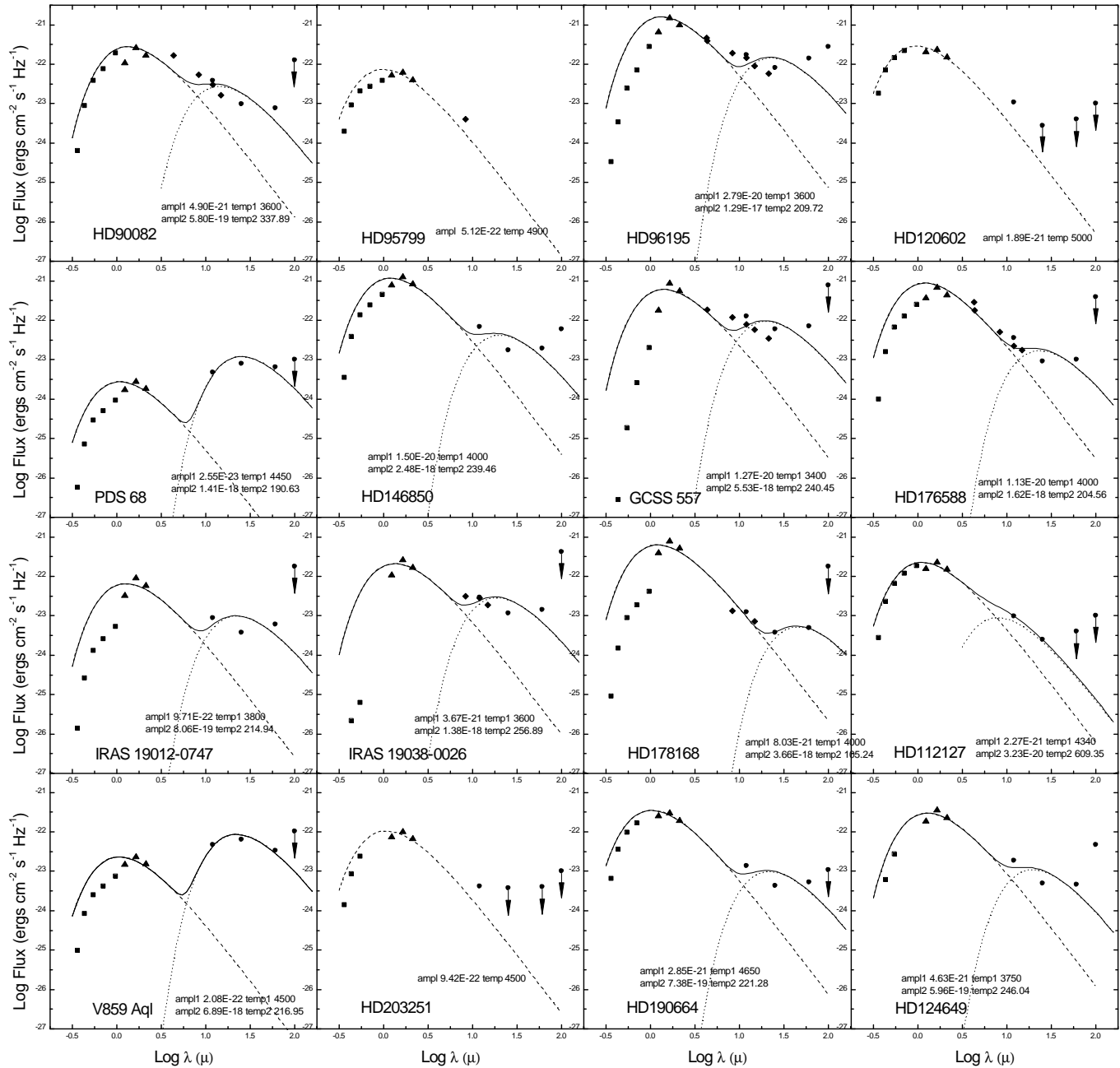}}
\caption{Spectral energy distributions. In squares, the optical data; in triangles, the NIR 2MASS data; in diamonds, the MSX data (band quality $ > $ 0); and in circles, tha IRAS data. In solid lines, the two black-body fit to the data. In dashed lines, the black-body with {\it T}$_{eff}$; and in dotted lines, the black body with the calculated dust temperature ({\it T}$_{dust}$).}
\label{seds}
\end{figure*}

\subsubsection{IRAS colors\label{irassec}}

In order to investigate if the large far infrared excess for LRG stars observed in IRAS colors is correlated with the optical intrinsic polarization, we identified each object of our sample with its IRAS counterpart (IRAS Point Source Catalogue). The results are shown in column (11) of Table \ref{nir}, and just one object (LRG: \object{HD 95779}) of our sample does not present an IRAS counterpart. 

Figure \ref{iras} shows the distribution of our sample in the IRAS 12-25-60 $ \mu $m color-color plane. The normal giants \object{HD190664} and \object{HD 124649} are coincident with the lower side of LRG stars. The locus of LRG stars in our sample is consistent with groups 1 and 3 from Gregorio-Hetem et al. (\cite{greg93}). Figure \ref{pir} plots the optical {\it P$_{int}$} versus [25 - 12] and [60 - 25] IRAS colors. Considering the objects with higher intrinsic polarizations (\object{GCSS 557} and \object{V859 Aql}), an apparent correlation is observed with [25-12] color and an anticorrelation with [60 -25] color. It could be consistent with our assumption of a favorable (edge-on) viewing angle for the envelope in these two objects. Again, the lower polarization observed in the rest of LRG stars would be reflecting objects with an intermediary viewing angle or pole-on envelopes. To quantity this finding, we calculated the {\it r}$_{s}$ coefficient (as in Sect. \ref{nirsec}) for all the data in Figs. \ref{pir}a,b. The calculated value of {\it r}$_{s}$ for the {\it P}$_{{\it int}}$ versus [25 - 12] color relation is 0.15 with a 70\% probability; and for the {\it P}$_{{\it int}}$ versus [60-25] color relation, {\it r}$_{s}$ is -0.72 with a 3\% probability. The positive (and less significant) and negative (and more significant) correlations with IRAS colors also can imply that dust with preferential emission in $ \sim $ 25 $\mu$m is responsible for the optical intrinsic polarization observed.


\begin{table} 
\caption{Additional information.}
\setlength\tabcolsep{10pt}
\begin{tabular}{llc}
\hline \hline 
Object     & MSX6C ID & {\it T}$_{dust}$ \\
     & &   (K) \\
    (1) & (2) & (3) \\
\hline
\object{HD 90082}	&	G286.2685-03.9091	&	338	\\
\object{HD 95799}	&	G289.1898+01.1718	&	--	\\
\object{HD 96195}	&	G291.1180-02.5694	&	210	\\
\object{HD 120602}	&	--	&	--	\\
\object{PDS 68}	&	--	&	191	\\
\object{HD 146850}	&	--	&	239	\\
\object{GCSS 557}	&	G017.0984-01.3757	&	240	\\
\object{HD 176588}	&	G030.0809-04.1814	&	205	\\
\object{IRAS 19012-0747}	&	--	&	215	\\
\object{IRAS 19038-0026}	&	G034.3214-03.4902	&	257	\\
\object{HD 178168}	&	G032.5404-04.7223	&	105	\\
\object{HD 112127}	&	--	&	609	\\
\object{V859 Aql}	&	--	&	217	\\
\object{HD 203251}	&	--	&	--	\\
\hline
\object{HD 190664}	&	--	&	221	\\
\object{HD 124649}	&	--	&	246	\\

\hline
\end{tabular}
\\
\label{tdust}
\end{table}

\subsection{{\it SED}s}

In order to check the typical dust temperature ({\it T}$_{dust}$) for the envelopes that are responsible for the optical polarization, we proceeded to construct the spectral energy distribution (SED) for the objects in our sample. For that we used the optical, near infrared, and far infrared data available (Tables \ref{log}, \ref{optical}, and \ref{nir}). We also used, when available, the intermediary infrared data from the MSX6C Infrared Point Source Catalog (Egan et al. \cite{egan03}). The objects with MSX data are indicated in column (2) of Table \ref{tdust}. The flux calibration was made using Cox (\cite{cox00}) for the optical data and Cohen et al. (\cite{cohe03}) for the NIR data. 

The results are shown in Fig. \ref{seds}. The optical and NIR data were not corrected by reddening. We also show an additive two black-body fit that includes a component with a fixed effective temperature (column (2) in Table \ref{optical}) representing the central object and a second component with the temperature as a free parameter representing {\it T}$_{dust}$ for the envelope. The fit was done over the NIR, MSX, and IRAS data to minimize the effect of reddening in the optical bands. The IRAS higher limits were not considered in the fit, while three objects (\object{HD 96195}, \object{HD 146850}, and \object{HD 124649}) with apparently good quality IRAS {\it F}$_{100}$ present cirrus contamination ({\it cirr}3/{\it F}$_{60}$ $ \ga$ (1- 5), Ivezic \& Elitzur \cite{ivez95}), and {\it F}$_{100}$ also were not considered. Typical dust temperatures for the envelopes were between 190 K and 260 K (see column (3) in Table \ref{tdust}) associated with maximum emission between 19 $ \mu $m and 14 $ \mu $m, respectively. This fact agrees with our conclusion about grains emitting in $ \sim $ 25 $\mu$m as responsible for the intrinsic polarizations (see Sect. \ref{irassec}).  As a comparison, Reddy et al. (\cite{redd02}) find {\it T}$_{dust}$ $\sim$ 250 K for V859 Aql, which compares with our value ({\it T}$_{dust}$ = 217 K) very well.

\section{Conclusions}

Optical polarimetry for a sample of fourteen LRG stars was obtained. For nine of them intrinsic polarization was estimated using field stars. Seven LRG (five of them with {\it P}/$\sigma$$_{{\it P}}$ $>$ 5) have lower but non-negligible intrinsic polarization (0.19 $-$ 0.34)\%, and in two cases (\object{V859 Aql} and \object{GCSS 557}) intrinsic polarizations higher than 0.5\% are found. These results indicate that an asymmetric spatial distribution of circumstellar dust is present in LRG (but probably also in normal giants). An excess in observed polarization when it is correlated with the optical excess color gives additional support to the circumstellar origin of the intrinsic polarization in LRG. The optical intrinsic polarization in LRG is not correlated with the near infrared excess but is correlated with far infrared emission. This would suggest that grains emitting in $\sim 25$ $\mu$m are responsible for the optical intrinsic polarization, and the higher intrinsic polarization levels would indicate a favorable (edge-on) viewing angle for the envelopes, as in \object{V859 Aql} and \object{GCSS 557}. Analysis of spectral energy distributions for the sample provides an estimate of the dust temperature for the envelopes, which are mainly between 190 K and 260 K. Our findings indicate that non-spherical symmetries may appear as early as the RG phases of stellar evolution.

\begin{acknowledgements}
The authors wish to thank the referee for his/her comments and suggestions that helped to improve this paper. A. Pereyra is thankful to CAPES and FAPESP (grant 02/12880-0) for financial support. A. M. Magalh\~aes acknowledges support from Fapesp and CNPq. Polarimetry at IAG-USP is supported by a FAPESP grant 01/12589-1. This research has made use of the SIMBAD database and VizieR catalogue access tool operated at CDS, Strasbourg, France. Also, this publication makes use of data products from the Two Micron All Sky Survey, which is a joint project of the University of Massachusetts and the Infrared Processing and Analysis Center/California Institute of Technology, funded by the National Aeronautics and Space Administration (NASA) and the National Science Foundation. Finally, this research made use of data products from the Midcourse Space Experiment.  Processing of the MSX data was funded by the Ballistic Missile Defense Organization with additional support from NASA Office of Space Science.  This research also made use of the NASA/ IPAC Infrared Science Archive, which is operated by the Jet Propulsion Laboratory, California Institute of Technology, under contract with the NASA.

\end{acknowledgements}



\end{document}